\documentclass[twocolumn,showpacs]{revtex4}
\usepackage{psfig}
\def\beq{\begin{equation}}
\def\eeq{\end{equation}}
\def\jp{J/\psi}
\def\pp{\psi'}

\def\gsim{\raisebox{-0.6ex}{$\stackrel{\textstyle >}{\sim}$}}
\begin{document}
\title{The interaction of slow $J/\psi$ and $\psi'$ with nucleons.}

\author{A. Sibirtsev$^{1,2}$ and M.B. Voloshin$^{3,4}$}

\affiliation{
$^1$Helmholtz-Institut f\"ur Strahlen- und Kernphysik (Theorie),
Universit\"at Bonn, Nu\ss allee 14-16, D-53115 Bonn, Germany \\
$^2$Institut f\"ur Kernphysik (Theorie), Forschungszentrum J\"ulich,
D-52425 J\"ulich, Germany,\\
$^3$William I. Fine Theoretical Physics Institute, University of
Minnesota, Minneapolis, MN 55455, USA \\
$^4$Institute of Theoretical and Experimental Physics, Moscow 117259, Russia}

\begin{abstract}
The interaction of the charmonium resonances $J/\psi$ and $\psi'$ with nucleons
at low  energies is considered using the multipole expansion and low-energy
theorems in QCD. A lower bound is established for the relevant gluonic operator
average over the nucleon. As a result we find the discussed interaction to be
significantly stronger than previously estimated in the literature. In
particular we conclude that the cross section of the $J/\psi$ - nucleon elastic
scattering at the threshold is very likely to exceed 17 millibarn and that
existence of bound states of the $J/\psi$ in light nuclei is possible. For the
$\psi'$ resonance we estimate even larger elastic scattering cross section and
also a very large cross section of the process $\psi'{+}N{\to}\psi{+}N$ giving
rise to the decay width of tens of MeV for the $\psi'$ resonance in heavy
nuclei.
\end{abstract}

\maketitle

\section{Introduction}
Understanding the charmonium interaction with nuclear matter is
important for description of the photo- and hadro- production of
charmonium and charmed hadrons on the nuclear targets as
well as for diagnostics of the hadronic final states in heavy-ion
collisions and search for Quark Gluon Plasma. Such interaction has
been a subject of numerous studies with a broad range of theoretical
predictions.

First perturbative QCD calculations~\cite{Peskin1,Bhanot} predicted  very small
$J/\psi$  dissociation cross section by  hadrons, on the order of few $\mu$barn.
Such small dissociation cross section supports the
prediction~\cite{Matsui,Kharzeev1,Kharzeev2} that a strong $J/\psi$ suppression
is generated by the QGP formation. Indeed, a strong $J/\psi$ suppression was
observed later in relativistic heavy ion collisions~\cite{Abreu}. On the
contrary, an alternative
explanation~\cite{Brodsky1,Armesto1,Capella,Sibirtsev1,Navarra} of the
suppression requires a substantial strength of the hadronic interaction of
$J/\psi$, corresponding to a scattering cross section on the order of few
millibarn. With all the great interest to the problem of charmonium interaction
with nucleons and nuclear matter and its practical importance, the discussion of
this interaction is still wide open. In particular,  the estimates of the
strength of the interaction of $\jp$ and $\pp$ with the nucleon range, in terms
of the scattering cross section at low energy, from a fraction of
millibarn~\cite{Kharzeev1,Kaidalov} up to 10~mb or
more~\cite{Brodsky2,Sibirtsev2}. Recent reviews of the subject and
further references can be found in the Refs.~\cite{Vogt,Barnes,Song}.

In many of these applications the most interesting energy region is
usually well above the threshold, where the complexity of the
problem becomes more confounding due to the multitude of possible
inelastic processes contributing to charmonium scattering on nuclear
matter. However the strength of the interaction at energy close to
the threshold is also measurable~\cite{Brodsky2}
and its reliable estimate can serve as an important reference point
for analyses of the behavior of the interaction at higher energies.
Furthermore, the $\jp$ and $\pp$ interactions at low energies are
of explicit importance for high energy heavy ion collisions since
the relative motion between the co-moving charmonium and nuclear
matter is rather small. In that case the charmonium scattering in hadronic gas
is given by the strength of the  $\jp$ and $\pp$ elastic interactions
and might be related to the transverse momentum component of the
charmonium spectra. Moreover the forward elastic scattering amplitude
can be related to  $\jp$ and $\pp$ mass shift in  matter predicted by
a number of models~\cite{Klingl,Hayashigaki,Kim,Tsushima1}.

Here we consider the interaction of the $\jp$ and $\pp$
resonances with nucleons and the nuclear matter at energies where the
least of inelastic processes are kinematically allowed. For the $\jp$
interaction with a nucleon in this region only the elastic scattering
contributes below the threshold of the lowest essential inelastic
charm-exchange channel $\jp{+}N{\to}\Lambda_c{+}{\bar D}$ at
$\sqrt{s}\, {\approx}4.15$~GeV. The subthreshold process
$\jp{+}N{\to}\eta_c{+}N$ should be significantly suppressed due
to the heavy quark spin-flip, and is entirely neglected in this study.
For the $\pp$ scattering, in addition to the elastic and the charm-exchange
processes, there always is the essential subthreshold process
$\pp{+}N{\to}\jp{+}N$, which is included in our discussion here.

The low-energy interaction of a heavy quarkonium with soft hadrons is
mediated by soft gluonic field, which should be treated nonperturbatively.
In this situation a use is made of the fact that the charmonium is
a relatively compact object in the scale set by $\Lambda_{QCD}$, so
that its interaction with soft gluons can be expanded in
multipoles~\cite{Gottfried,Voloshin1}. The leading term in this
expansion arises from the double $E1$ interaction with the chromo-electric
component of the gluonic field, and the heavy quarkonium part of this
interaction can be parametrized in terms of the quarkonium
chromo-polarizability. The coupling of the soft gluonic fields to
light hadrons, specifically to the nucleons, at a low momentum transfer
is determined by the low-energy theorem in QCD based on the anomaly
in the trace of the energy-momentum tensor. An application of such
approach\cite{Voloshin2,Novikov1} to hadronic transitions in charmonium
$\pp{\to}\jp\pi\pi$ and bottomonium $\Upsilon^\prime{\to}\Upsilon\pi\pi$
is known to be in a good agreement with the data~\cite{Voloshin3}. Here we find
a lower bound for the relevant gluonic matrix element over nucleon and we also
argue that its actual value should be close to this bound.

The paper is organized as follows. In Sec.2 we discuss the chromo-polarizability
of the charmonium states, which arises within the multipole expansion in QCD
used for description of the properties of heavy quarkonium states and
transitions between them. In Sec.3 we relate the discussed scattering amplitudes
to the chromo-polarizability and to the matrix element of the square of the
chromo-electric field over nucleon. The latter matrix element is described by
the low-energy theorem following from the conformal anomaly in QCD. In Sec.4 the
threshold limit of the $\jp$ - nucleon scattering amplitude is considered and
estimated in terms of the scattering length, the cross section, and the average
potential energy of $\jp$ in nuclear matter. The same characteristics are
considered in Sec.5 for the $\pp$ - nucleon interaction, with the addition of
the inelastic subthreshold process $\pp{+}N{\to}\jp{+}N$, which also contributes
to decay of the $\pp$ in nuclear matter. Finally, Sec.6 contains our concluding
remarks.

\section{Charmonium chromo-polarizability}

The leading $E1$ term in the multipole expansion for the interaction
of a heavy quarkonium with soft gluon field has the
form~\cite{Gottfried,Voloshin1}
\beq
H_{E1}=-{1 \over 2} \xi^a \, {\vec r} \cdot {\vec E}^a (0)~,
\label{e1}
\eeq
where $\xi^a{=}t_1^a{-}t_2^a$ is the difference of the color generators
acting on the quark and antiquark, e.g. $t_1^a{=}\lambda^a/2$ with
$\lambda^a$ being the Gell-Mann matrices, and ${\vec r}$ is the vector
for relative position of the quark and the antiquark. We use here the
normalization for the gluon field where the QCD coupling $g$ is
included in the definition of the field, so that e.g. the gluon field
Lagrangean reads as $L{=}{-}(F_{\mu \nu}^a)^2/(4 g^2)$. The amplitude
of the transition between $S$-wave states $A$ and $B$ of the heavy
quarkonium in the second order in $H_{E1}$ can then be written in
terms of the effective operator
\beq
\langle B | H_{eff}| A \rangle =-{1 \over 2} \, \alpha_{AB} \,
{\vec E}^a \cdot {\vec E}^a~,
\label{heff}
\eeq
where the non-relativistic normalization is used for the quarkonium
states, and $\alpha_{AB}$ is the chromo-polarizability, which can be found as
\beq
\alpha_{AB}={1 \over 48} \,\langle B |\, \xi^a \, r_i \, G_A \, r_i \, \xi^a
| A \rangle ~.
\label{dme}
\eeq
Here $G_A$ is the Green's function for a heavy quark pair in
color-octet (adjoint) state. This Green's function is not well
understood presently, therefore a theoretical calculation of the
chromo-polarizability is at least highly model-dependent.

The value of the chromo-polarizability for the transition
$\pp{\to}\jp$, namely $\alpha_{\pp \jp}$, determines the amplitude of the
decay $\pp{\to}\jp\pi\pi$~\cite{Voloshin2}, and can thus be
found\cite{Voloshin4} from the known decay rate:
$|\alpha_{\pp \jp}|{\approx}2~{\rm GeV}^{-3}$.

The diagonal values $\alpha_{\jp}$ and $\alpha_{\pp}$ are not known
presently. These can be measured experimentally in the
decays $\jp{\to}\ell^+\ell^-\pi^+\pi^-$ and
$\pp{\to}\ell^+\ell^-\pi^+\pi^-$ with soft pions\cite{Voloshin4}.
It is however natural to expect that each of the diagonal amplitudes
should be somewhat larger than the transition amplitude. Since the
polarizability grows with the spatial size of the system, it is
also natural to expect that this parameter is larger for
$\pp$ than for the $\jp$, $|\alpha_{\pp}|{>}|\alpha_{\jp}|$.
The general restrictions on the diagonal amplitudes arise from the
fact that in the Green's function in eq.(\ref{dme}) there are no
intermediate states with mass below $\pp$, so that both
$\alpha_{\jp}$ and  $\alpha_{\pp}$ are real and positive, and their
values satisfy the inequality:
\beq
\alpha_{\pp} \, \alpha_{\jp} \ge |\alpha_{\pp \jp}|^2~.
\label{schw}
\eeq
Naturally, this inequality implies that at least one of the diagonal
amplitudes is larger than the transition one, although this statement
is of course weaker than the natural expectation that each of the
discussed diagonal chromo-polarizabilities exceeds the known value
of $|\alpha_{\pp \jp}|$.

Given these estimates, we use the value $2~{\rm GeV}^{-3}$ as a
reference for the discussed parameters $\alpha_{\jp}$ and
$\alpha_{\pp}$, keeping in mind that their actual values can
be somewhat larger, especially that of $\alpha_{\pp}$.

It should be noted that our `reference' value significantly exceeds
the estimate of $\alpha_{\jp}$ by Kaidalov and Volkovitsky\cite{Kaidalov}.
In that estimate they used the approach~\cite{Bhanot} based on essentially
a Coulomb-like model for charmonium wave functions, and their result
can be written as
$\alpha_{\jp} = {28 \over 81}\, \pi a^3$,
where $a$ is the size parameter for charmonium (the Bohr radius
in the Coulomb-like model), for which they used $a{=}0.8~{\rm GeV}^{-1}$.
Numerically, their estimate corresponds to
$\alpha_{\jp}{\approx}0.6~{\rm GeV}^{-3}$, which we believe is too
low, given the arguments presented above and the known value of the
transition amplitude $\alpha_{\pp \jp}$.

\section{The nucleon matrix element}
The effective operator in eq.(\ref{heff}) can be directly used for
calculating the amplitude of the scattering of the heavy quarkonium
on a nucleon $A+N{\to}B+N$ in terms of the matrix element of the
gluon operator ${\vec E}^a{\cdot}{\vec E}^a$ over the nucleon:
\beq
T_{AB}=2 \sqrt{M_A \, M_B} \, \alpha_{AB} \, \langle N|
{1 \over 2}\, {\vec E}^a \cdot {\vec E}^a |N \rangle~,
\label{tab}
\eeq
where the factor $2\sqrt{M_A \, M_B}$ appears due to
the relativistic normalization of the scattering amplitude $T$,
which normalization is used in the rest of this paper in order to facilitate
direct comparison with Ref.\cite{Kaidalov}.

The matrix element over the nucleon can be evaluated following
the approach\cite{Novikov1} used for estimating similar matrix element
over pions. Namely, one writes
\begin{eqnarray}
{1 \over 2}\, {\vec E}^a \cdot {\vec E}^a={1 \over 4}\,
\left ( {\vec E}^a \cdot {\vec E}^a - {\vec B}^a \cdot
{\vec B}^a \right) \nonumber \\
+ {1 \over 4}\, \left ( {\vec E}^a
\cdot {\vec E}^a + {\vec B}^a \cdot {\vec B}^a \right) \nonumber \\
=-{1 \over 8}\, (F_{\mu \nu}^a)^2 + 2 \pi \, \alpha_s \, \theta^{00}_G~,
\label{dec}
\end{eqnarray}
where $\theta^{\mu \nu}_G$ is the energy-momentum tensor of the
gluon field, and relates the term with $(F_{\mu \nu}^a)^2$ to the
expression for the anomalous trace of the full energy-momentum
tensor in QCD in the chiral limit:
\beq
-{b \over 32 \pi^2} \, (F_{\mu \nu}^a)^2 = \theta^\mu_\mu~,
\label{anom}
\eeq
with $b{=}9$ being the first coefficient in the QCD beta function
with three light (massless in the chiral limit) quarks. Using
these relations the matrix element over the nucleon entering
eq.(\ref{tab}) can be written as
\beq
\langle N| {1 \over 2}\, {\vec E}^a \cdot {\vec E}^a |N
\rangle={4 \pi^2 \over b} \langle N| \theta^\mu_\mu |N
\rangle +  2 \pi \, \alpha_s \, \langle N |\theta^{00}_G|N
\rangle~.
\label{me2}
\eeq
The first term in the latter expression at a small momentum
transfer $q{=}p_1-p_2$ can be found using
\beq
{4 \pi^2 \over b}\, \langle N(p_2)| \theta^\mu_\mu
|N(p_1) \rangle={4 \pi^2 \over b}\, m_N \,{\bar N}(p_2) N(p_1)~.
\label{trmn}
\eeq
The only approximation made here is in neglecting the difference
between the actual mass of the nucleon $m_N$ and its value in
the chiral limit. Although the correction for this difference
can be taken into account, we neglect it, since this
correction is likely to be less than other uncertainties in our estimates.

The last term in eq.(\ref{me2}) is formally of higher order
in the QCD coupling $\alpha_s$, and it has been neglected in
Ref.\cite{Kaidalov} in comparison with the anomalous contribution.
We argue here however, that in the amplitude under discussion the
contribution of this term is at least as large as that of the
anomalous one. Indeed, consider the discussed matrix element at
zero momentum transfer $q$ for the nucleon being at rest. The
equation (\ref{anom}) can then be written as the diagonal
average of the gluonic operator
\beq
\langle N| {1 \over 4}\, \left ( {\vec E}^a \cdot {\vec E}^a
- {\vec B}^a \cdot {\vec B}^a \right)|N \rangle = {4 \pi^2
\over b}\, 2 m_N^2~.
\label{astat}
\eeq
On the other hand the diagonal average (over a non-vacuum state)
of the manifestly quadratic operator ${\vec B}^a{\cdot}{\vec B}^a$
has to be non-negative. Thus one arrives at the inequality
\beq
\langle N| {1 \over 2}\, {\vec E}^a \cdot {\vec E}^a |N \rangle
\ge {8 \pi^2 \over b} \, 2 m_N^2~,
\label{ineq}
\eeq
which corresponds to that for a static nucleon the second term
in the final expression in eq.(\ref{me2}) is  at least as large
as the first term. It can be noticed in connection with
the discussed bound that these two terms are of the same order
in $N_c$ in the large $N_c$ counting. It is reasonable to expect
that the chromo-magnetic average over the nucleon, although
non-negative, is substantially smaller than the chromo-electric
one, so that the actual chromo-electric amplitude is close to the
lower bound required by eq.(\ref{ineq}).

It is also instructive to analyze the matrix element in
eq.(\ref{me2}) within the approach used by Novikov and
Shifman~\cite{Novikov1} for an estimate of a similar amplitude
for dipion production by gluonic field in the transition
$\pp{\to}\jp\pi\pi$, where it has been found that the contribution
of the gluonic part of the energy-momentum tensor is indeed small
in comparison with that of the anomalous term in the relevant
kinematics. According to this approach the matrix element of the
gluonic part of the energy-momentum tensor is parametrized in
terms of the fraction $\rho_G$ of the nucleon energy and momentum
carried by gluons, and the discussed term can be written as
\begin{eqnarray}
2 \pi \, \alpha_s \, \langle N(p_2) |\theta^{00}_G|N(p_1)
\rangle{=} \nonumber \\
\pi \, \alpha_s \rho_G \, (p_1^0 + p_2^0) \, N^\dagger(p_2) N(p_1)~,
\label{ener}
\end{eqnarray}
where $p_1^0$ ($p_2^0$) is the energy of the initial (final)
nucleon. The appropriate normalization scale for the product
$\alpha_s \rho_G$ is the characteristic size of the heavy
quarkonium. Novikov and Shifman estimate for the case of
the dipion transition in charmonium $\alpha_s \rho_G{\approx}0.7$.
By writing the expression (\ref{ener}) in the static limit, one finds that
in the case of the nucleon amplitude discussed here, the
inequality (\ref{ineq}) in fact requires the relevant
parameter $\alpha_s \rho_G$ for the nucleon to be at least as
large as estimated in Ref.\cite{Novikov1} for the pions:
$\alpha_s \rho_G{\ge}0.7$.

The discussed here low-energy static limit for the nucleon matrix
element in the amplitude $T$ in eq.(\ref{tab}) is sufficient
for considering the threshold limit of the elastic $\jp$-nucleon
interaction. However in the subthreshold process
$\pp{+}N{\to}\jp+N$ with slow $\pp$ the momentum transfer is
quite substantial:
$-q^2{=}(M_{\pp}^2-M_{\jp}^2)\, m_N/(M_{\pp}+m_N){\approx}0.82~{\rm GeV}^2$,
where $M_{\jp}$ ($M_{\pp}$) stands for the mass of the $\jp$ ($\pp$)
resonance. The form factor, describing the deviation of the discussed
nucleon matrix element from its value at $q^2=0$ is presently unknown.
However, for the case of the similar matrix element over pions it
is known from the study of the shape of the dipion spectrum in the
transition $\Upsilon'{\to}\Upsilon\pi\pi$ that this deviation is rather
small. Namely, if the form factor $F(q^2)$ at small $q^2$ is parametrized
as $F{=}1{+}q^2/M^2+\ldots$, the coefficient of the first term
corresponds to $M{>}1~{\rm GeV}$ (a discussion of this topic can
be found in the review~\cite{Voloshin3}). The relatively
large mass scale in this form factor agrees well with the
general arguments~\cite{Novikov2} for the presence of a large
mass scale in the $0^{++}$ flavor-singlet hadronic channel. In
what follows, we use the simple parametrization of the nucleon
matrix element in terms of its value at $q^2{=}0$ and one common form factor:
\begin{eqnarray}
\langle N(p_2)| {1 \over 2}\, {\vec E}^a \cdot {\vec E}^a |N(p_1) \rangle
={ 4 \pi^2 \over b} \, \left [ m_N \, {\bar N}(p_2) N(p_1) \right.
\nonumber \\ \left. + C {p_1^0 + p_2^0 \over 2} \, N^\dagger(p_2) N(p_1)
\right ] \, F(q^2)~,
\label{param}
\end{eqnarray}
where the constant $C$ describes the second term in
eq.(\ref{me2}) relative to the first one, and is bound by the
inequality (\ref{ineq}) as $C{\ge}1$. The form factor is normalized
as $F(0){=}1$, which point is relevant for the low-energy
$\jp$-nucleon scattering and for the real part of the
elastic $\pp$-nucleon scattering amplitude, while for
the subthreshold process $\pp+N{\to}\jp+N$ one inevitably has
to make assumptions about the behavior of the form factor.

\section{The $\jp$-nucleon interaction at the threshold}
The expression (\ref{tab}) can now be applied to evaluating
the threshold amplitude of the $\jp$ - nucleon scattering:
\beq
T_{\jp}= {16 \pi^2 \over 9} (1+C) \,  \alpha_{\jp} \,  M_{\jp} \, m_N^2~.
\label{tjp}
\eeq
The corresponding scattering length is then found as
\begin{eqnarray}
a_{\jp}={T_{\jp} \over 8 \pi \, (M_{\jp}+m_N)} \nonumber \\
\approx 0.37~{\rm fm}
\times  \left[ \frac{1+C}{2}
\frac{\alpha_{\jp}}{2 \, {\rm GeV}^{-3}} \right],
\label{sljp}
\end{eqnarray}
where the term in brackets indicates potential uncertainties due to
our present knowledge of the parameters $C$ and $\alpha_{\jp}$.
Accordingly, the $\jp N$ elastic scattering cross section is estimated as
\beq
\sigma_{\jp N} = 4 \pi \, a_{\jp}^2 \approx 17 \, {\rm mb} \times \left[
\frac{1+C}{2}
\frac{\alpha_{\jp}}{2 \, {\rm GeV}^{-3}} \right]^2.
\label{sjp}
\eeq
We find that our estimate of the scattering length exceeds the
previous one~\cite{Kaidalov} by at least seven times, and
consequently the estimated cross section is also at least 50
times larger.

Within the low density theorem the result  for
the elastic scattering amplitude can be converted into an
estimate of the  $\jp$ potential
in nuclear matter with
the density of nucleons $\rho_N{=}0.16~{\rm fm}^{-3}$:
\beq
V_{\jp}=- \frac{T_{\jp}\, \rho_N}{4 \, M_{\jp}\,  m_N} \approx -21 \,
{\rm MeV} \times \left[ \frac{1+C}{2}
\frac{\alpha_{\jp}}{2 \, {\rm GeV}^{-3}} \right],
\label{vjp}
\eeq
which can be compared with the previous estimates given in Table~\ref{tab1}.
Note that in our notation the $\jp$ mass shift $\Delta M_{\jp}$ equals to
the real part of the potential~\cite{Hayashigaki,Lenz,Dover}. Table~\ref{tab1}
also shows the elastic $\jp{+}N$ cross section. Moreover we do not indicate
the elastic cross section evaluated~\cite{Sibirtsev2,Redlich} from
$\jp$ photoproduction off proton, since there are no available data
close to the threshold~\cite{Sibirtsev3}.

\begin{table}[t]
\caption{The mass shift $\Delta M_{\jp}$ and elastic $\jp{+}N$
cross section predicted by different models.}
\label{tab1}
\vspace{1.5mm}
\begin{ruledtabular}
\begin{tabular}{lcc}
Ref. &  $-\Delta M_{\jp}$  (MeV) & $\sigma_{\jp N}$ (mb) \\
\colrule
\cite{Kaidalov} & 3 & 0.3 \\
\cite{Sibirtsev2} & & 1.5 \\
\cite{Klingl}   & 10$\div$5 & \\
\cite{Hayashigaki} & 7$\div$4 & \\
\cite{Kim} & 4 & \\
\cite{Luke} & 11${\div}$8 & \\
\cite{Teramond} & & 5 \\
\cite{Brodsky3} & & 8 \\
\cite{Lee} & 5 &   \\
this  & \gsim \, 21 & \gsim \, 17 \\
\end{tabular}
\end{ruledtabular}
\end{table}

The significant increase in the estimated potential also changes
the conclusion about the possibility of existence of  bound states
of $\jp$ in light nuclei. Indeed, the condition for existence of a
bound state in the approximation, where a nucleus is considered as
being of a uniform density $\rho_N$ up to the sharp boundary at the
radius $R_A$ reads as
\beq
R_A^2 > {\pi^2 \over 8 M_{\jp} (-V_{\jp})}~.
\label{ra}
\eeq
With the minimal estimate of the binding potential in eq.(\ref{vjp})
this condition is satisfied already at $R_A{>}0.9\,{\rm fm}$, which points to a
relevance of the problem of bound states to light nuclei.
Although the criterion in eq.(\ref{ra}) is not directly applicable for light
nuclei, the resulting estimate gives credibility to the claims
\cite{Brodsky2,Wasson} that bound states of the $\jp$ resonance
in nuclei do exist starting from light nuclei.

With regards to existence of a near-threshold bound or resonant
state of the $\jp$ and a single nucleon, the presented here
consideration is generally insufficient for arriving at a
definite conclusion. It should be noted, however, that if such
near-threshold singularity exists, it would require a substantial
modification of the presented here estimates for the scattering
amplitude. The estimate of the amplitude in eq.(\ref{tjp}) by itself,
although larger than previously thought, generally does not
require a unitary modification. Indeed, the estimated amplitude
is still below the unitarity limit as long as the c.m. momentum
satisfies the condition $p{<}a_{\jp}^{-1}{\approx}530~{\rm MeV}$,
so that at such momenta the unitarity corrections are relatively small.

\section{The $\pp$ interaction with nucleons}
The formulas of the previous section for the interaction of a
slow $\jp$ with nucleons and nuclei can be directly applied to
calculation of the real part of the $\pp$-nucleon scattering
amplitude, and of the energy shift of the $\pp$ in nuclear matter,
by the obvious replacement
$M_{\jp}{\to}M_{\pp}$ and $\alpha_{\jp}{\to}\alpha_{\pp}$.
In interpreting the resulting formulas in numerical terms one
should keep in mind that the expected chromo-polarizability
$\alpha_{\pp}$ is very likely to substantially exceed the `reference'
value $2~{\rm GeV}^{-3}$. In particular this implies that
the binding energy of the $\pp$ in heavy nuclei,
\beq
V_{\pp} \approx -21 \,
{\rm MeV} \times \left[ \frac{1+C}{2}
\frac{\alpha_{\pp}}{2 \, {\rm GeV}^{-3}} \right],
\label{vpp}
\eeq
should be
significantly larger than the numerical value 21 MeV.
This shift in the energy of the $\pp$ can be important for the estimates
of the decay $\pp{\to}D{\bar D}$, which becomes possible in nuclear
matter due to the shifts in the masses of the $D$ and ${\bar D}$
mesons~\cite{Tsushima2,Sibirtsev4,Friman,Tsushima3,Mishra}.

The main difference between the nuclear interactions of
slow $\jp$ and $\pp$ is that for the latter there exist
subthreshold scattering processes: the charm-exchange process
$\pp{+}N{\to}\Lambda_c{+}{\bar D}$, the charmonium transition
scattering $\pp{+}N{\to}\jp{+}N$, and generally additional channels
where in the latter process instead of a single nucleon excited
states are being produced such as $N\pi$, $\Delta\pi$, etc. The
processes other than $\pp{+}N{\to}\jp{+}N$ are beyond the scope
of the present paper. We can only comment here that due to the
discussed relation of the relevant gluonic matrix element to
the energy-momentum tensor in QCD, the processes with non-diagonal
transitions, such as $N{\to}N\pi$, should be suppressed with
respect to the diagonal one $N{\to}N$. It can be also noted
that similar transitions from $\pp$ to lower charmonium states
other than $\jp$ should also be suppressed in comparison with
$\pp{\to}\jp$, since those other states cannot be produced in
the second order in the leading $E1$ term of the multipole expansion.

The scattering amplitude for the process $\pp{+}N{\to}\jp{+}N$
with slow $\pp$ is found from the equations (\ref{tab}) and
(\ref{param}), where in the latter equation the momentum transfer
is fixed at $-q^2{\approx}0.82~{\rm GeV}^2$. Clearly, at such
momentum the form factor $F(q^2)$ can be significantly different
from one. The present poor knowledge of this form factor results
in the largest uncertainty in estimating the scattering amplitude.
In view of this uncertainty we simplify the rest of the matrix
element in eq.(\ref{param}) by neglecting the kinetic energy
of the final nucleon, thus writing the scattering amplitude in the form
\beq
T_{\pp \jp} \approx {16 \pi^2 \over 9} (1+C) \,
\alpha_{\pp \jp} \, \sqrt{M_{\jp} M_{\pp}} \, m_N^2 \, F(q^2)~.
\label{tpj}
\eeq
Using this amplitude one readily finds the scattering cross
section near the $\pp{+}N$ threshold at the c.m. momentum $p_i$
of the initial particles:
\begin{eqnarray}
\sigma(\pp+N{\to}\jp+N)={1 \over p_i} {|T_{\pp \jp}|^2 \,
p_f \over 16 \pi \, (M_{\pp}+m_N)^2 } \nonumber \\
\approx 16 \, {\rm mb} \, \left[ {1 \, {\rm GeV}
\over p_i} \right] \, \left[ {1+C \over 2} \right]^2 \, |F(q^2)|^2~,
\label{spj}
\end{eqnarray}
where $p_f \approx 1.0 \, {\rm GeV}$ is the c.m. momentum
in the final state. The inverse-velocity, $1/p_i$, behavior
of the cross section is due the subthreshold kinematics of
the process. Assuming, conservatively, that the form factor
$|F(q^2)|$ suppresses the amplitude by not more than a factor
of two, one comes to the conclusion that the cross section of
the considered process can reach tens of millibarn at rather
moderately low values of the initial momentum $p_i$.

Furthermore, the unitarity relation implies that the amplitude
of the considered inelastic process contributes to the imaginary
part of the amplitude $T_{\pp}$ of the elastic $\pp{+}N$ scattering
near threshold:
\beq
{\rm Im} \, T_{\pp}=|T_{\pp \jp}|^2 {p_f \over 8 \pi (M_{\pp}+m_N)}~.
\label{uni}
\eeq
Using the formula in eq.(\ref{tab}) for the real part of $T_{\pp}$
and the approximation in eq.(\ref{tpj}) for the amplitude
$T_{\pp \jp}$ one arrive at the following estimate of the
significance of this effect in terms of the ratio of the
imaginary to the real part of the elastic scattering amplitude:
\begin{eqnarray}
{{\rm Im} \, T_{\pp} \over {\rm Re} \, T_{\pp}}
\approx {2 \pi \over 9} \, (1+C) \, {|\alpha_{\pp \jp}|^2
\over \alpha_{\pp}} \, {M_{\jp} \, m_N^2 \, p_f \over M_{\pp}+m_N} \,
|F(q^2)|^2
\nonumber \\ \approx 1.6 \, \left[ {1+C \over 2} \right] \,
\left[ 2 \, {\rm GeV}^{-3} \over \alpha_{\pp} \right]
\, |F(q^2)|^2~.
\label{imre}
\end{eqnarray}
Given the expected form factor suppression, and that, as discussed,
$\alpha_{\pp}$ should be larger than $2~{\rm GeV}^{-3}$, one can
conclude from this estimate that the contribution of the discussed
inelastic channel to the imaginary part of the elastic scattering
amplitude is still somewhat smaller than the real part.

The effect of the discussed imaginary part of the amplitude,
although relatively moderate in the scattering cross section,
gives rise to a potentially interesting effect when considered
in terms of the imaginary part of the average potential energy
$V_{\pp}$ of $\pp$ in nuclear medium with the nucleon density
$\rho_N$. The imaginary part of the binding energy corresponds
to the decay rate $\Gamma_{\pp}{=}-2~{\rm Im} \, V_{\pp}$,
and the contribution of the process $\pp{+}N{\to}\jp{+}N$ to
$\Gamma_{\pp}$ can be directly estimated from eq.(\ref{tpj}) as
\begin{eqnarray}
\Gamma_{\pp\jp}=|T_{\pp \jp}|^2 {p_f \over 32 \pi (M_{\pp}+m_N)\, M_{\pp} \,
m_N} \,
\rho_N \nonumber \\
\approx 70 \, {\rm MeV} \, \left [ {1+C \over 2} \right ]^2 \,
\left [\alpha_{\pp}  \over 2 \, {\rm GeV}^{-3} \right ]^2 \, |F(q^2)|^2~,
\label{ppg}
\end{eqnarray}
and is likely reaching tens of MeV at the average nuclear
density $\rho_N{\approx}0.16~{\rm fm}^{-3}$.

\begin{table}[t]
\caption{The mass shift $\Delta M_{\pp}$ predicted
by different models.}
\label{tab2}
\vspace{1.5mm}
\begin{ruledtabular}
\begin{tabular}{lc}
Ref. &  $-\Delta M_{\jp}$  (MeV) \\
\colrule
\cite{Luke} & $700$ \\
\cite{Lee} & $130$ \\
this  &  $> 21$  \\
\end{tabular}
\end{ruledtabular}
\end{table}

\section{Concluding remarks}
The approach used here to calculation of the interaction of the
$\jp$ and $\pp$ resonances with nucleons is based on the notion
that the heavy quarkonium
is a compact object and its interaction with soft gluon field can
be expanded in multipoles. For the discussed processes such
approximation should work best for the elastic $\jp$ - nucleon
interaction at energy close to the threshold. For this reason
we believe that the estimates for this process suffer from the
least uncertainty. The largest uncertainty in the numerical
estimates for this case arises from the presently unknown
chromo-polarizability of the $\jp$ state of charmonium.
This parameter however can be measured in the decay
$\jp{\to} \ell^+\ell^-\pi^+\pi^-$, thus eliminating the largest
source of uncertainty. The other unknown involved in our estimates
is the coefficient $C$ for the ratio of the non-anomalous to the anomalous part
of the gluonic matrix element in eq.(\ref{me2}).
The inequality (\ref{ineq}) bounds the value of this coefficient
as $C{\ge}1$, and it can be reasonably argued that the actual
value should be close to this bound. However at present we
cannot suggest a way of an independent determination of
this coefficient.

The accuracy of the considered approach
becomes worse with increasing energy in the $\jp$ - nucleon
system, since the gluonic fields mediating the interaction
become less soft, thus worsening the applicability of the
multipole expansion. For this reason it is troublesome at
present to interpolate between our estimates in the near-threshold
region and other theoretical approaches to the interaction at
higher energies. For both charmonium resonances we find
the scattering cross section on a nucleon
at threshold to be quite large:
$17~{\rm mb}$ or more for the $\jp$, and in the tens of millibarn
for the $\pp$. A comparison of our estimates with the only available
experimental value\cite{Anderson} for the $\jp$ - nucleon total cross
section: $\sigma_{\jp N}{=}3.8{\pm}0.8{\pm}{\sim}0.5~{\rm mb}$
at $\sqrt{s}{\approx}5.7~{\rm GeV}$ suggests a
noticeable rise of the cross section toward the threshold.
At present, however, we are not aware of any
arguments that would exclude a considerable decrease of the cross
section away from the threshold.

The issue of applicability of the multipole expansion is still
more sensitive for the case of the $\pp$ resonance, which naturally
has larger characteristic size than the $\jp$ state. An application
of the expansion in this case is to a certain extent
justified by the very good agreement of the data on the
charmonium dipion transition $\pp{\to}\jp\pi\pi$ with the description
based on the same approach. However the departure of the
interaction from the considered low-energy limit at higher
energies should be more rapid than for the case of $\jp$.
The signature of such departure is in fact relevant already in
the threshold limit for the inelastic process $\pp{+}N{\to}\jp{+}N$
where the momentum transfer is non-negligible. In the presented
here calculation the effect of the deviation from the strictly
static limit is encoded in the form factor $F(q^2)$, which
inevitably adds to the uncertainty of the presented estimates.

We believe that even with a rather conservative assumption about
this form factor at the actual value of the momentum transfer, one
can conclude that the inelastic process gives a large contribution
to the decay width of the $\pp$ resonance in heavy nuclei.
Furthermore, it should be noted that the process $\pp{+}N{\to}\jp{+}N$
is not the only one contributing to the decay. Another potentially
large contribution can come from the charm-transfer process
$\pp{+}N{\to}\Lambda_c{+}{\bar D}$, which is entirely different
from the type of processes considered in this paper, hence
we do not present here any further discussion of this process.

The substantial modification of $\jp$ and $\pp$ properties in nuclear
medium can be studied experimentally. It is a
challenge of the future Facility for Antiproton and Ion Research (FAIR) at
GSI~\cite{Schwarz,Ritman,Senger1,Senger2} to provide
valuable data for further progress in understanding the
QCD dynamics in matter.

\section*{Acknowledgments}
One of us (MBV) acknowledges the hospitality of the Institut
f\"ur Kernphysik, Forschungszentrum J\"ulich, where most of
this work was done during the visit supported by an award from
the Alexander von Humboldt Foundation.

The work of MBV is supported, in part, by the DOE grant DE-FG02-94ER40823.


\begin{thebibliography}{99}
\bibitem{Peskin1}
        M.E. Peskin, Nucl. Phys. B {\bf 156}, 365 (1979).
\bibitem{Bhanot}
        G. Bhanot and M.E. Peskin, Nucl. Phys. B {\bf 156}, 391 (1979).
\bibitem{Matsui}
         T. Matsui and H. Satz, Phys. Lett. B  {\bf 178}, 416 (1986).
\bibitem{Kharzeev1}
         D. Kharzeev and H. Satz, Phys.Lett. B {\bf 334}, 155 (1994).
\bibitem{Kharzeev2}
         D. Kharzeev, Nucl. Phys. A {\bf 638}, 279c (1998).
\bibitem{Abreu}
        M.C. Abreu et al., Phys. Lett. b {\bf 477}, 28 (2000)
\bibitem{Brodsky1}
        S.J. Brodsky and A.H. Mueller, Phys. Lett. B {\bf 206}, 685 (1988)
\bibitem{Armesto1}
        N. Armesto and A. Capella, J. Phys. G {\bf 23}, 1969 (1997).
\bibitem{Capella}
        A. Capella, E.G. Ferreiro and A.B. Kaidalov, Phys. Rev. Lett.
        {\bf 85}, 2080 (2000).
\bibitem{Sibirtsev1}
        A. Sibirtsev, K. Tsushima, K. Saito and  A.W. Thomas,
        Phys. Lett. B {\bf 484}, 23 (2000).
\bibitem{Navarra}
        F.S. Navarra, M. Nielsen, R.S. Marques de Carvalho, G. Krein,
        Phys. Lett. B {\bf 529}, 87 (2002).
\bibitem{Kaidalov}
        A.B. Kaidalov and P.E. Volkovitsky, Phys.Rev.Lett.
        {\bf 69}, 3155 (1992).
\bibitem{Brodsky2}
        S.J. Brodsky, I. Schmidt, and G.F. de Terramond,
        Phys. Rev. Lett. {\bf 64}, 1011 (1990).
\bibitem{Huefner}
        J. H\"ufner {\it et.al.}, Phys. Rev. D {\bf 62}, 094022 (2000).
\bibitem{Sibirtsev2}
        A. Sibirtsev, K. Tsushima, and A.W. Thomas, Phys. Rev. C
        {\bf 63}, 044906 (2001).
\bibitem{Vogt}
         R. Vogt, Phys. Rept. {\bf 310}, 197 (1999).
\bibitem{Barnes}
        T. Barnes,  Eur. Phys. J. A {\bf 18}, 531 (2003).
\bibitem{Song}
        T. Song and S.H. Lee, hep-ph/0501252
\bibitem{Klingl}
        F. Klingl, S. Kim, S.H Lee, P. Morath and W Weise, Phys. Rev.
        Lett. {\bf 82} 3396 (1999), Erratum-ibid. {\bf 83}, 4224 (1999).
\bibitem{Hayashigaki}
        A. Hayashigaki, Prog. Theor. Phys. {\bf 101}, 923 (1999).
\bibitem{Kim}
        S. Kim and  S.H. Lee, Nucl. Phys. A {\bf 679}, 517 (2001).
\bibitem{Tsushima1}
        K. Tsushima, A. Sibirtsev, K. Saito, A.W. Thomas and D.H. Lu,
        Nucl. Phys. A {\bf 680}, 280 (2001).
\bibitem{Gottfried}
        K. Gottfried, Phys. Rev.Lett. {\bf 40}, 538 (1978).
\bibitem{Voloshin1}
        M.B. Voloshin, Nucl. Phys. B {\bf 154}, 365 (1979).
\bibitem{Voloshin2}
        M.B. Voloshin and V.I. Zakharov, Phys. Rev. Lett. {\bf 45}, 688 (1980).
\bibitem{Novikov1}
        V.A. Novikov and M.A. Shifman, Z. Phys. C {\bf 8}, 43 (1981).
\bibitem{Voloshin3}
        M.B. Voloshin and Yu.M. Zaitsev, Usp. Fiz. Nauk, {\bf 152},
        361 (1987);\ [Sov. Phys. Usp. {\bf 30}, 553 (1987)].
\bibitem{Voloshin4}
        M.B. Voloshin, Mod.Phys.Lett. A {\bf 19}, 665 (2004).
\bibitem{Novikov2}
        V.A. Novikov {\it et.al.}, Nucl. Phys. B {\bf 191}, 301 (1981).
\bibitem{Lenz}
        W. Lenz, Z. Phys. {\bf 56}, 778 (1929).
\bibitem{Dover}
        C.D. Dover, J. H\"ufner and R.H. Lemmer, Ann. Phys. {\bf 66},
        248 (1971).
\bibitem{Luke}
         M.E. Luke, A.V. Manohar, M.J. Savage, Phys. Lett. B
         {\bf 288}, 355 (1992).
\bibitem{Teramond}
        G.F. de Teramond, R. Espinoza and M. Ortega-Rodrigues,
        Phys. Rev. D {\bf 58}, 034012 (1998)
\bibitem{Brodsky3}
        S.J. Brodsky and A.H. Mueller, Phys. Lett. B {\bf 412}, 125 (1997)
\bibitem{Lee}
        S.H. Lee and C.M. Ko, Phys. Rev. C {\bf 67}, 038202 (2003)
\bibitem{Redlich}
        K. Redlich, H. Satz and  G.M. Zinovjev, Eur. Phys. J. C
        {\bf 17}, 461 (2000)
\bibitem{Sibirtsev3}
        A. Sibirtsev, S. Krewald and  A.W. Thomas, J. Phys. G
        {\bf 30}, 1427 (2004)
\bibitem{Wasson}
        D.A. Wasson, Phys. Rev. Lett. {\bf 67}, 2237 (1991)
\bibitem{Tsushima2}
        K. Tsushima, D.H. Lu, A.W. Thomas, K. Saito and R.H. Landau,
        Phys. Rev. C {\bf 59}, 2824 (1999).
\bibitem{Sibirtsev4}
        A. Sibirtsev, K. Tsushima and A.W. Thomas,  Eur. Phys. J.
        A {\bf 6}, 351 (1999).
\bibitem{Friman}
        B. Friman, S.H. Lee and  T. Song, Phys. Lett.
        B {\bf 548}, 153 (2002)
\bibitem{Tsushima3}
        K. Tsushima and  F.C. Khanna (Adelaide U.), Phys. Lett. B
        {\bf 552}, 138 (2003).
\bibitem{Mishra}
        A. Mishra, E.L. Bratkovskaya, J. Schaffner-Bielich, S. Schramm
        and  H. Stocker, Phys. Rev. C {\bf 69}, 015202 (2004).
\bibitem{Sibirtsev5}
        A. Sibirtsev, Eur. Phys. J. A {\bf 18}, 475 (2003).
\bibitem{Anderson}
        R.L. Anderson {\it et.al.}, Phys.Rev.Lett. {\bf 38}, 263 (1977).
\bibitem{Schwarz}
        C. Schwarz et al., Phys. Scripta T {\bf 104}, 147 (2003).
\bibitem{Ritman}
        J. Ritman, Eur. Phys. J. A {\bf 18}, 177 (2003).
\bibitem{Senger1}
        P. Senger, J. Phys. G {\bf 30}, S1087 (2004).
\bibitem{Senger2}
        P. Senger, Acta Phys. Polon. B {\bf 35}, 1131 (2004).
\end{thebibliography}
\end{document}